\documentclass[prb,twocolumn]{revtex4-1}
\usepackage[utf8]{inputenc}
\usepackage{amsmath,amsfonts,amssymb,siunitx}
\usepackage{graphicx,subfigure}
\usepackage{booktabs}

\begin{document}
\title{Effective Hamiltonian for two-electron Quantum Dots \\
from weak to strong parabolic confinement}
\author{Torsten Victor Zache}
\author{Aniruddha Chakraborty}
\affiliation{School of Basic Sciences, Indian Institute of Technology, Mandi 175001}
\date{\today}

\begin{abstract}
We model quasi-two-dimensional two-electron Quantum Dots in a parabolic confinement potential with rovibrational and purely vibrational effective Hamiltonian operators. These are optimized by non-linear least-square fits to the exact energy levels.  We find, that the vibrational Hamiltonian describes the energy levels well and reveals how relative contributions change on varying the confinement strength. The rovibrational model suggests the formation of a rigid two-electron molecule in weak confinement and we further present a modified model, that allows a very accurate transition from weak to strong confinement regimes.
\end{abstract}
\maketitle

\section{Introduction}
Since the discovery of Quantum Dots (hereafter QDs) in 1981\cite{ekimov1981quantum}, a lot of research has been carried out in the direction of semiconductor nanostructures. Few-electron\cite{johnson1995quantum} QDs, especially two-electron QDs are of huge theoretical interest, because they provide a model system for studying electron-electron correlation \cite{ihn2007quantum} and properties of electronic spectra in general (e.g. Hund's rule \cite{sako2012origin}). Experiments\cite{heitmann1993spectroscopy} suggest, that the confinement can be modeled by a parabolic potential as a good approximation in many cases. In this paper we only examine quasi-two-dimensional two-electron QDs in the absence of external fields. For external magnetic fields, the reader is referred to other work.\cite{maksym1990quantum,chakraborty1992physics} Nevertheless the energy levels in two-electron QD exhibit a complicated behaviour. In the strong confinement regime, electron correlation effects can be neglected to a good degree of approximation \cite{bryant1987electronic} and the spectrum exhibits some degeneracies as a result of the generealized Kohn theorem \cite{resonance1961cyclotron,dobson1994harmonic}. For weak confinement the two electrons behave similar to a rigid molecule with characteristical rotational and vibrational energy spectra \cite{yannouleas2000collective}.

In this paper, we present possible unified descriptions of two-electron Quantum Dots from weak to strong confinement regimes. Section 2 gives different approaches to construct an effective Hamiltonian operator for the energy spectrum. One is based on the rovibrational "molecular" model and the other one is purely vibrational using three coupled Harmonic Oscillators. Several non-linear fits are performed in order to improve the accuracy of the operators. We give our results for both weak and strong confinement in Section 3 and interpret them in terms of electronic correlation. Section 4 summarizes and discusses our findings.

\section{Rovibrational and vibrational effective Hamiltonian}
We consider two electrons in a QD with parabolic confinement potential. The exact Hamiltonian (using effective atomic units) reads
\begin{equation}
H = - \nabla_1^2 - \nabla_2^2 + \frac{1}{4}\omega_0^2 r_1^2 + \frac{1}{4}\omega_0^2 r_2^2 + \frac{2}{|\vec{r}_1 - \vec{r}_2|}
\label{eq:exactH}
\end{equation}
Numerically the energy levels can be calculated exactly \cite{zhu1996exact} and we use these for fitting our effective energy operators. The data we have used is given in TAB. \ref{tab:data}. Varying the confinement strength $\omega_0$ basically changes the relative importance of the confining and the Coulomb repulsion potential. We will use this to test the validity of our models and to interpret the effect of electron-electron correlation. 

\begin{figure}[t]
\caption{\label{fig:rovib}Selected states from TAB. \ref{tab:data} showing a molecularlike spectrum. (a) equally spaced vibrational excitations; (b) rotational excitations}
\includegraphics[scale=0.40]{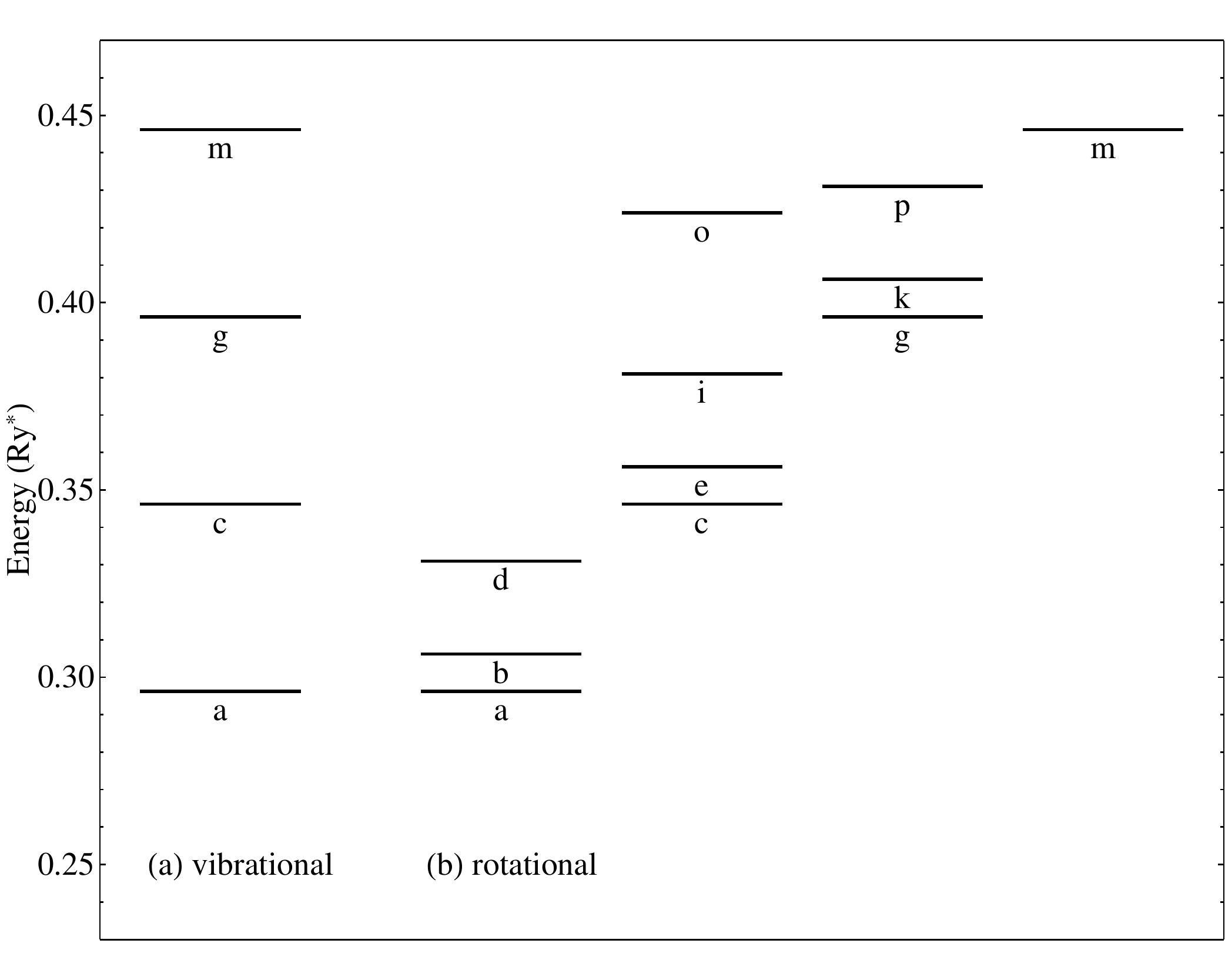}
\end{figure}

\begin{table}[b]
\caption{\label{tab:data}Exact energy levels for \eqref{eq:exactH} given in $\SI{}{Ry^*}$. The labelings correspond to the rovibrational and purely vibrational effective Hamiltonian operators. The missing labels are discussed at the end of section 3. Quantum numbers for centre-of-mass, relative-motion and spin can be found in the reference paper.\cite{zhu1996exact}}
\begin{ruledtabular}
\begin{tabular}{ccccc}
	& $(N,M)$	& $(n_0,n_a,n_b)$	& $\omega_0 = 1.0$	& $\omega_0 = 0.05$\\
\hline
a:	& (0,0)	& (0,0,0)		& (a) 3.3196			& (a) 0.2962\\
b: 	& (0,1) 	& (0,0,1)		& (b) 3.8278			& (b) 0.3062\\
c:	& (1,0) 	& (1,0,0)		& (c) 4.3196			& (d) 0.3310\\
d:	& (0,2) 	& (0,1,1)		& (d) 4.6436			& (c) 0.3462\\
e:	& (1,1)	& (1,0,1)		& (e) 4.8278			& (h) 0.3476\\
f:	&		&			& (f) 5.1472			& (e) 0.3562\\
g:	& (2,0)	& (2,0,0)		& (g) 5.3196			& (i) 0.3810\\
h:	&		&			& (h) 5.5174			& (f) 0.3854\\
i:	& (1,2)	& (1,1,1)		& (i) 5.6436			& (g) 0.3854\\
j:	& 		&			& (j) 5.7438			& (j) 0.3968\\
k:	& (2,1)	& (2,0,1)		& (k) 5.8278			& (k) 0.4062\\
l:	& 		&			& (l) 6.1472			& (n) 0.4066\\
m: 	& (3,0)	& (3,0,0)		& (m) 6.3196			& (o) 0.4240\\
n:	&		& 			& (n) 6.4693			& (p) 0.4310\\
o:	& (1,3)	& (1,1,2)		& (o) 6.5956			& (l) 0.4354\\
p:	& (2,2)	& (2,1,1)		& (p) 6.6436			& (m) 0.4462
\end{tabular}
\end{ruledtabular}
\end{table}
In the case of weak confinement the energy spectrum exhibits a molecular-like picture consisting of rotational and vibrational contributions as can be seen in FIG. \ref{fig:rovib}. 
Therefore it seems reasonable to use a standard effective Hamiltonian of the following form (we have ignored an irrelevant constant)
\begin{equation}
H_{eff}^{(1)} = w \left( N+\frac{1}{2} \right) + BM^2 -DM^4
\label{eq:rovib}
\end{equation}
Here the vibrational $(N)$ and angular momentum $(M)$ quantum numbers have been assigned in a natural way (see TAB. \ref{tab:data}). \eqref{eq:rovib} is known to describe molecular spectra like the one shown in FIG. \ref{fig:rovib} very well, even with $D=0$, but the latter correction will be necessary for stronger confinement. We also want to mention that a similar formula has been derived analytically from the exact Hamiltonian.\cite{puente2004roto} Note that we are dealing with a two-dimensional system, so that $M^2$ should be used instead of the three-dimensional $J(J+1)$ term. However, we also consider another form leading to surprising results.
\begin{equation}
\tag{$2'$}
H_{eff}^{(1')} = w \left( N+\frac{1}{2} \right) + BM(M+1) -DM^2{(M+1)}^2
\label{eq:rovib'}
\end{equation}

Schwinger \cite{schwinger} pointed out, that it is possible to describe angular momentum with two uncoupled harmonic oscillators. Since correction terms for angular momentum are necessary here, we use coupled oscillators instead. Empirically we arrive at the following  purely vibrational effective Hamiltonian
\begingroup
\small
\begin{eqnarray}
H_{eff}^{(2)} = w_1 \left( n_0+\frac{1}{2} \right)+w_2 \left( n_a+n_b+1 \right) \label{eq:vib}\\*
+\alpha \left( \left( n_a+\frac{1}{2} \right) {\left( n_b + \frac{1}{2} \right)}^2 + {\left( n_a + \frac{1}{2} \right)}^2  \left( n_b+\frac{1}{2} \right) \right) & \nonumber
\end{eqnarray}
\endgroup
In this case, the labels have been chosen, so that the relations
\begin{subequations}
\begin{eqnarray}
n_0 =& N\\
n_a+n_b =& M\\
|n_a-n_b| =&  \text{min.} \label{eq:condition}
\end{eqnarray}
\end{subequations}
are fulfilled. The condition \eqref{eq:condition} allows to describe the states with the given effective Hamiltonian \eqref{eq:vib} by coupling the different oscillator modes. The parameter $\alpha$ determines the strength of the coupling between the "angular" vibrations $n_a$ and $n_b$.

\section{Results and Interpretation}
\begin{figure*}
\caption{\label{fig:results}Results of the fits for weak and strong confinement on left and right side, respectively; (a) exact data \cite{zhu1996exact}; (b) rovibrational model \eqref{eq:rovib}; (c) vibrational model \eqref{eq:vib}; (d) modified rovibrational model \eqref{eq:rovib'}; the labels are explained in TAB. \ref{tab:data}}
\subfigure{
\includegraphics[scale=0.495]{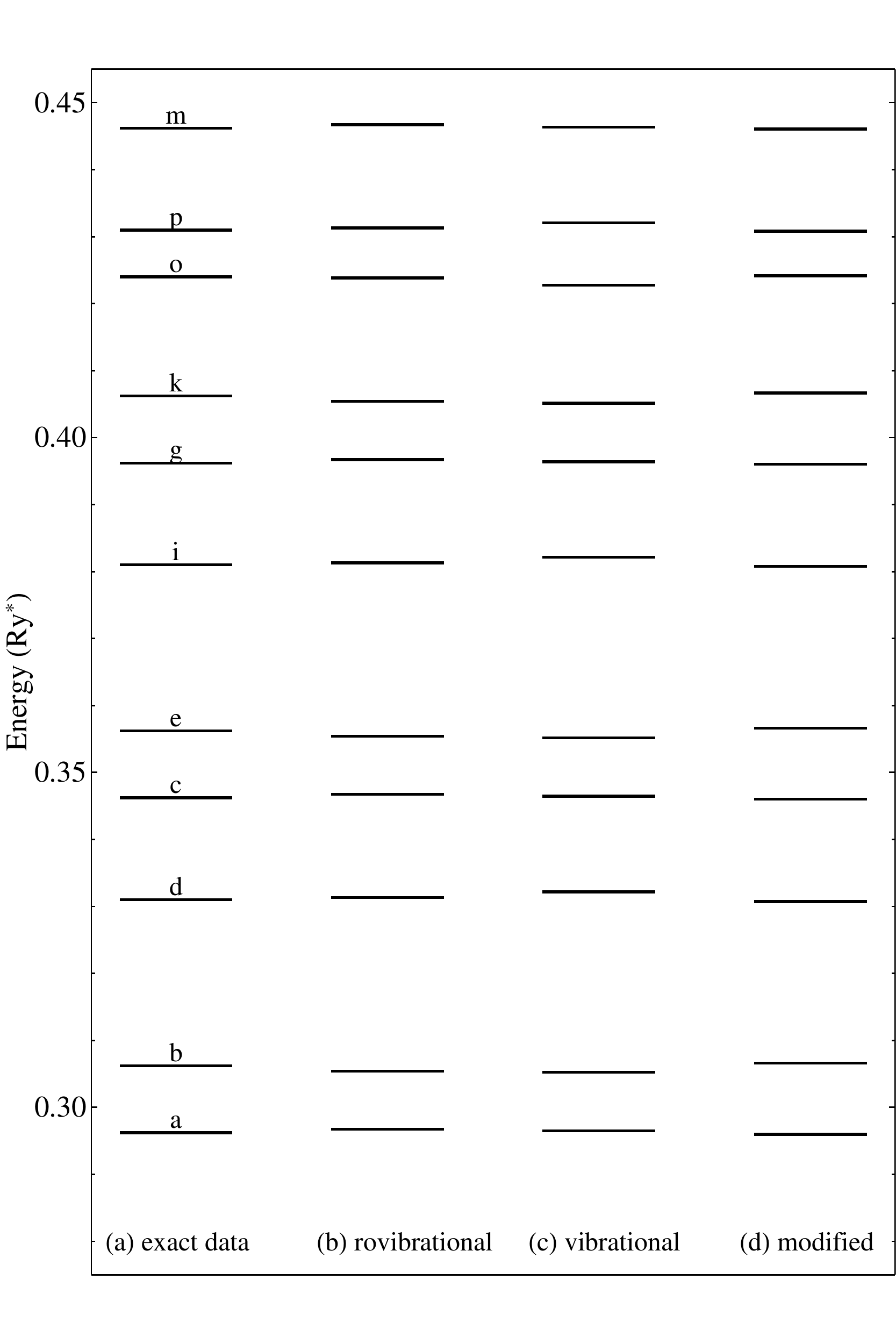}}
\hfill
\subfigure{
\includegraphics[scale=0.49]{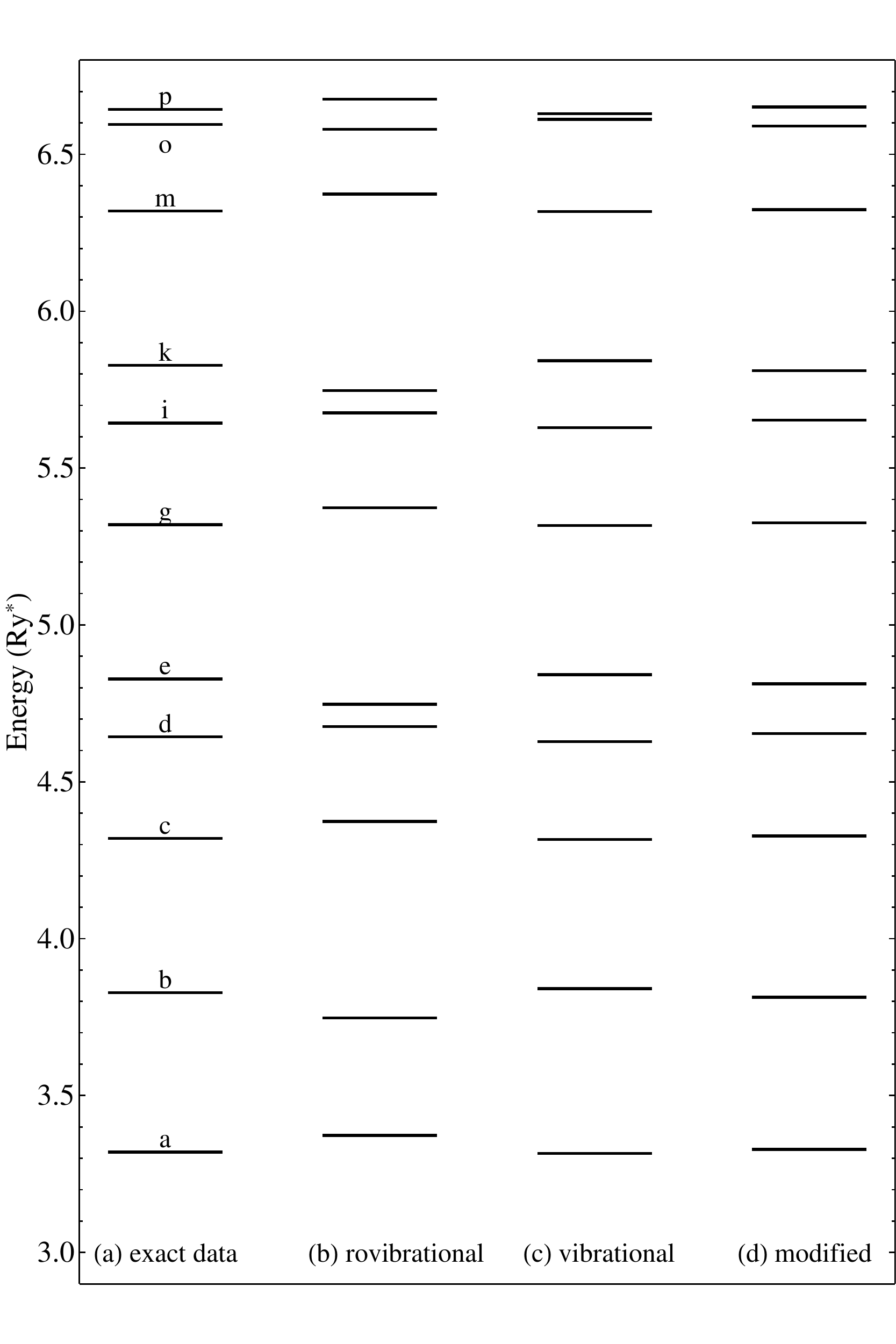}}
\end{figure*}

We have fitted the energy levels in TAB. \ref{tab:data} to the operators \eqref{eq:rovib} and \eqref{eq:vib} by means of non-linear least-square fits in order to optimze their parameters. Each fit consists of 11 independent levels and is used to determine 3 parameters. The results are given in TAB. \ref{tab:results}.
\begin{table}[t]
\caption{\label{tab:results}Results of the fits for weak ($\omega_0 = \SI{0.05}{Ry^*}$) and strong ($\omega_0 = \SI{1.0}{Ry^*}$) confinement given in units of $\SI{e-3}{Ry^*}$ and $\SI{}{Ry^*}$, respectively.}
\begin{ruledtabular}
\begin{tabular}{ccccccc} 
				&\multicolumn{3}{c}{$\omega_0 = 50.0$}   					& \multicolumn{3}{c}{$\omega_0 =1.0$} \\
				& $H^{(1)}$ 	& $H^{(1')}$ 	& $H^{(2)}$ 	& $H^{(1)}$ 	& $H^{(1')}$ 	& $H^{(2)}$\\
\hline
$\omega, \omega_1$ 	& 50.0		& 50.0		& 49.9		& 1.000		& 0.999		& 1.001\\
$B$				& 8.7			& 5.1			& -			& 0.390		& 0.253		& -\\
$D$				& 0.0			& -0.1			& -			& 0.016		& 0.005		& -\\
$\omega_2$			& - 			& -			& 3.0			& - 			& - 			& 0.443\\
$\alpha$			& -			& - 			& 4.6			& - 			& - 			& 0.065\\
rms				& 1.7			& 0.9			& 3.0			& 0.177		& 0.035		& 0.039
\end{tabular}
\end{ruledtabular}
\end{table}  
Both models fit the energy levels reasonably well, as can be seen from the low rms values. The rovibrational model \eqref{eq:rovib} is more accurate than the vibrational model \eqref{eq:vib} in the case of weak confinement, which is not very surprising, because the spectrum obviously shows a rovibrational structure (see FIG. \ref{fig:rovib}). But for strong confinement it does not describe the energy levels as well as the vibrational model. We also performed fits for the modified rovibrational model \eqref{eq:rovib'} yielding surprisingly low rms values in both confinement regimes. The results are also pictured in FIG. \ref{fig:results}. We want to emphazise, that all effective Hamiltonians correctly describe the level-crossing of the different states. In the rovibrational picture, this crossing can be explained by a broadening of the single rotational spectra (compare FIG. \ref{fig:rovib} and FIG. \ref{fig:results}). 

The calculated values of $\omega, \omega_1$ coincide with the confinement strength $\omega_0$ as they are expected to. We find, that $D=0$ for the simple rovibrational model in case of weak confinement, where electron repulsion is strong enough to fix the distance between the two electrons and leads to the formation of a rigid two-electron molecule. For stronger confinement $D \neq 0$, so that the rigidity becomes less accurate. $D$ can be thought of as centrifugal distortion - a common interpretation in molecular spectroscopy. In this semi-classical picture, the stronger confinement forces the electrons to stay closer together and increases their effective moment of inertia. 

At first sight, the modified model looks very similar to the rovibrational one. Although we also call it "rovibrational", one has to be very careful about this interpretation. The modified model describes a three-dimensional rotor-vibrator, which is not expected here. Another fit for this Hamiltonian with fixed $D=0$ for weak confinement reveals a rms value of 5.6, which proves that the "centrifugal distortion" term is needed to correct the predictions of the model, even in the small regime. We also want to mention, that the sign of $D$ changes from weak to strong confinement, while the relative contribution (as compared to $B$) stays the same. The change in sign makes a classical interpretation as centrifugal distortion rather difficult. Still both models have a smaller angular contribution for weak confinement and a stronger angular contribution for strong confinement in common, as can be seen by comparing the relevant values of $B$ (and $D$) to $\omega$.

The purely vibrational Hamiltonian \eqref{eq:vib} is not as accurate as the modified rovibrational model, but its components make this operator even more intersting. The angular contribution is implemented by two harmonic oscillator modes $n_a$ and $n_b$ and a term describing a coupling between them. The fact, that the operator is invariant under exchange of $n_a$ and $n_b$ suggests, that electron correlation might be partly described by these oscillators. Closer examination of the coupling term supports an interpretation as contribution to electon correlation energy. Indeed, this part  looses relative importance (again comparing the calculated parameters) in the large regime. This is in agreement with the expectation, that electron correlation has less influence in strong confinement. However, it cannot be neglected completely.

\begin{figure}[t]
\caption{\label{fig:contrib}Average relative contributions of the parameters of the effective vibrational Hamiltonian \eqref{eq:vib} normalized to unity. The weak and strong confinement regimes are shown in green (light grey) and black, respectively.}
\includegraphics[scale=0.32]{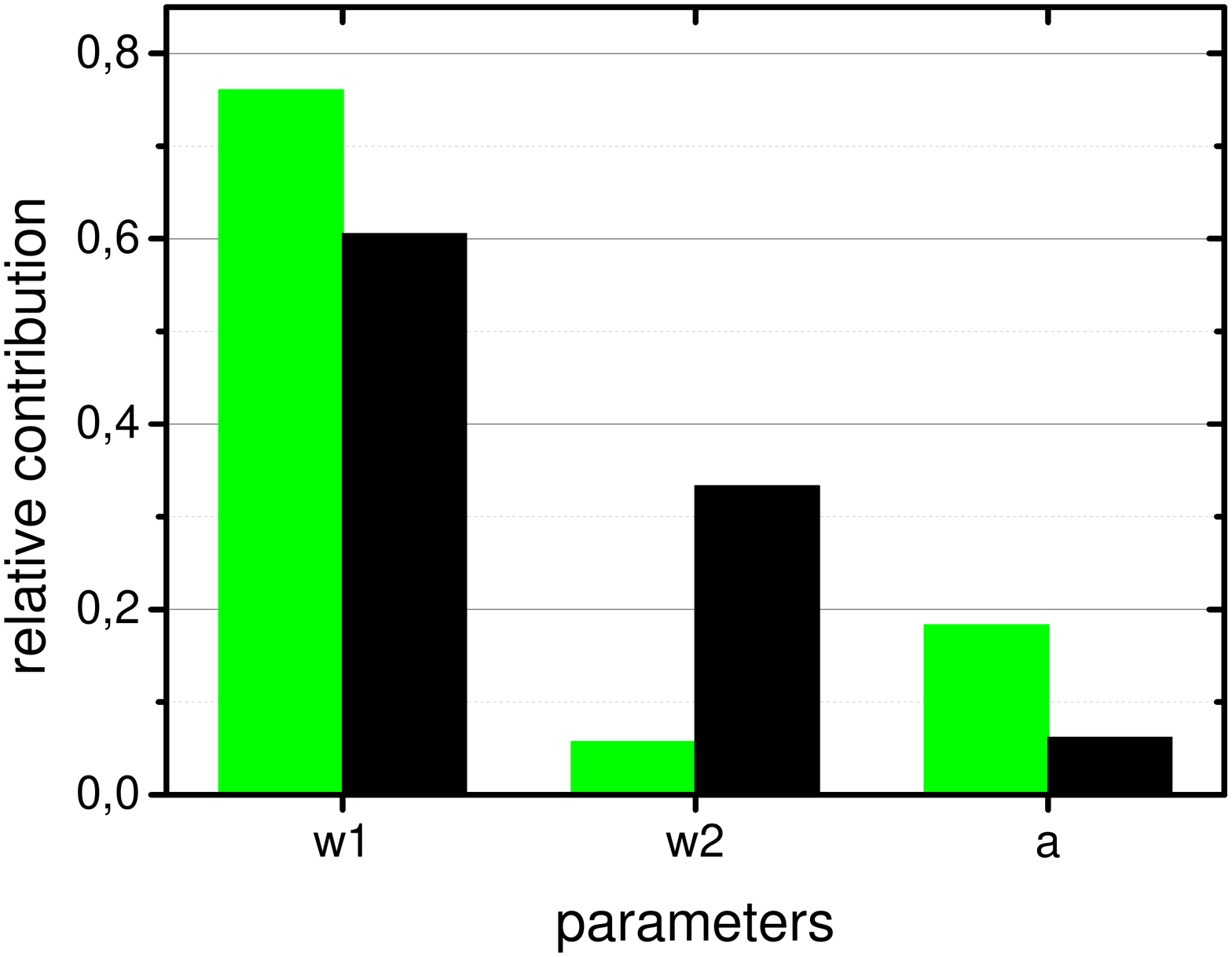}
\end{figure}

The strongest contribution stems from the "main" vibration $n_0$, but it decreases for increasing confinement and the "angular" vibrations $n_{a/b}$ gain importance, just like the angular contribution in the rovibrational model. To show the effect of the confinement strength on the different components of our model we have plotted their relative contributions in FIG. \ref{fig:contrib}.
Since the different terms enter with very different strengths depending on the considered state, we have calculated average contributions. Still one has to be careful interpreting FIG. \ref{fig:contrib}. Consider the state o: (1,1,2) as an extreme example. It is easy to calculate, that the relative contribution of the "angular" vibrations (the $w_2$ term) increases by a factor of 7 from weak to strong confinement. The fact, that it becomes even stronger than the "main" vibrational contribution (the $w_1$ term) explains the occurance of level crossings in the vibrational picture.  

Examining FIG. \ref{fig:contrib}, one might think, that the coupling term does not play an important role at all. Another fit with fixed $\alpha = 0$ reveals the rms values 23.9 and 0.343 for weak and strong confinement, respectively. This proves the importance of the corresponding term in \eqref{eq:vib}. However, it also shows, that the coupling term contains more information than only electron correlation. Otherwise it could approximately be neglected in strong confinement.

We want to give some additional thoughts about the labeling of states for our models. Note, that we have been very careful with the use of the expression "quantum number". Since the corresponding operators  do not commute with the given effective Hamiltonians, we think it is more appropriate to speak of approximate quantum numbers or just labels. However, our assignment for the rovibrational model is in good agreement with the center-of-mass and relative-motion description.\cite{zhu1996exact} At least for the lower rotational levels, $M$ coincides with the angular quantum number of relative-motion. States with odd and even $M$ are singlet and triplet, respectively. The careful reader might have noticed, that the energy of the state l (in weak confinement) suggests, that it belongs to the first rotational spectrum. However, we did not choose to label it (0,4), because it is a singlet, not a triplet and the state (0,3) is missing in the data. We note, that the average energy of f and h (in weak confinement) coincides with the expected (0,3) energy. This could hint at a splitting of these two states, which remains to be explained. We also mention, that the states n and l exhibit near degeneracies with k and p, which are broken in stronger confinement. Some of these observations could probably be described by additional vibrational modes, if the electrons are considered to form a "triatomic" molecule (with the center-of-mass as the third component). However, the data considered here is not sufficient to test this hypothesis.

\section{Summary and Conclusion}
In this paper, we have constructed different effective Hamiltonians for quasi-two-dimensional two-electron QDs in parabolic confinement. The rovibrational model strongly supports the formation of a rigid two-electron molecule in the weak confinement regime. We presented another modified rovibrational model, that describes the energy spectrum surprisingly well for both weak and strong confinement. The third model also allows a very accurate description in the large regime and is a little bit less useful for weak confinement. However, it is of great interest, because it is purely vibrational. Two coupled harmonic oscillators are used to replace the angular contribution of the rovibrational model allowing a transition from weak to strong confinement.

In conclusion, the presented effective Hamiltonians allow an accurate description of the energy spectrum of two-electron QDs in weak and strong confinement regimes. It will be interesting to test the validity of the models in even stronger confinement regimes and for a larger amount of excited states. The explanation of the surprisingly high accuracy of the modified rovibrational model remains the subject of further research. Furthermore, we want to point out, that the vibrational model could be used in more complicated situations, like three-electron QDs, where a "molecular" description as the rigid rotor-vibrator has not been found yet. We strongly believe, that the examination of few-electron QDs with a purely vibrational description will lead to a deeper understanding of electron correlation in general.

\begin{acknowledgements}
Some of the figures for this article have been created using the LevelScheme scientific figure
preparation system [M. A. Caprio, Comput. Phys. Commun. 171, 107 (2005),
http://scidraw.nd.edu/levelscheme].

We would like to thank the German Academic Exchange Service (DAAD) for providing scholarship for Torsten Zache during his internship at IIT Mandi.
\end{acknowledgements}

\bibliography{references}

\end{document}